\def\ba{\begin{eqnarray}}

\def\ea{\end{eqnarray}}
\def\lab{\label}
\def\n{\nonumber \\}
\def\b{\bibitem}

\documentclass[12pt]{article}
\usepackage{epsfig}
 \begin{document} 
\title{Bremsstrahlung from colour charges as a source of soft particle production 
in hadronic collisions}

\author{A.Bialas  and M.Jezabek\\ M.Smoluchowski Institute of Physics \\Jagellonian
University, Cracow\thanks{Address: Reymonta 4, 30-059 Krakow, Poland;
e-mail:bialas@th.if.uj.edu.pl;}\\ Institute of
Nuclear Physics PAN,  Cracow\thanks{Address: Radzikowskiego 152, 31-342 Krakow, Poland}
\\ AGH - University of Science and Technology, Cracow\thanks{Address: Faculty of Physics and 
Nuclear Techniques AGH, Kawiory 26a, 30-055 Krakow, Poland }}
\maketitle



\begin{abstract} It is proposed that soft particle production in
hadronic collisions is dominated by multiple gluon exchanges between
partons from the colliding hadrons, followed by radiation of hadronic
clusters from the coloured partons distributed uniformly in rapidity.
This explains naturally two dominant features of the data: (a) The
linear increase of rapidity spectra in the regions of limiting
fragmentation and, (b) the proportionality between the increasing width
of the limiting fragmentation region and the height of the central
plateau.

\end{abstract}

\newpage 
{\bf 1.} Recently, a substantial evidence is accumulating that particle
production in hadron-hadron, hadron-nucleus and nucleus-nucleus
collisions at high energy satisfies the principle of limiting
fragmentation \cite{lf} in a much wider range of rapidities than
originally proposed. This effect, first seen in $p\bar{p}$ \cite{ua5} and
 in p-Emulsion collisions \cite{h}, was recently studied in $d-Au$
and $Au-Au$ interactions by the PHOBOS collaboration \cite{ph1,ph2}. 

%rys.1
\begin{figure}[htb]
\centerline{
\epsfig{file=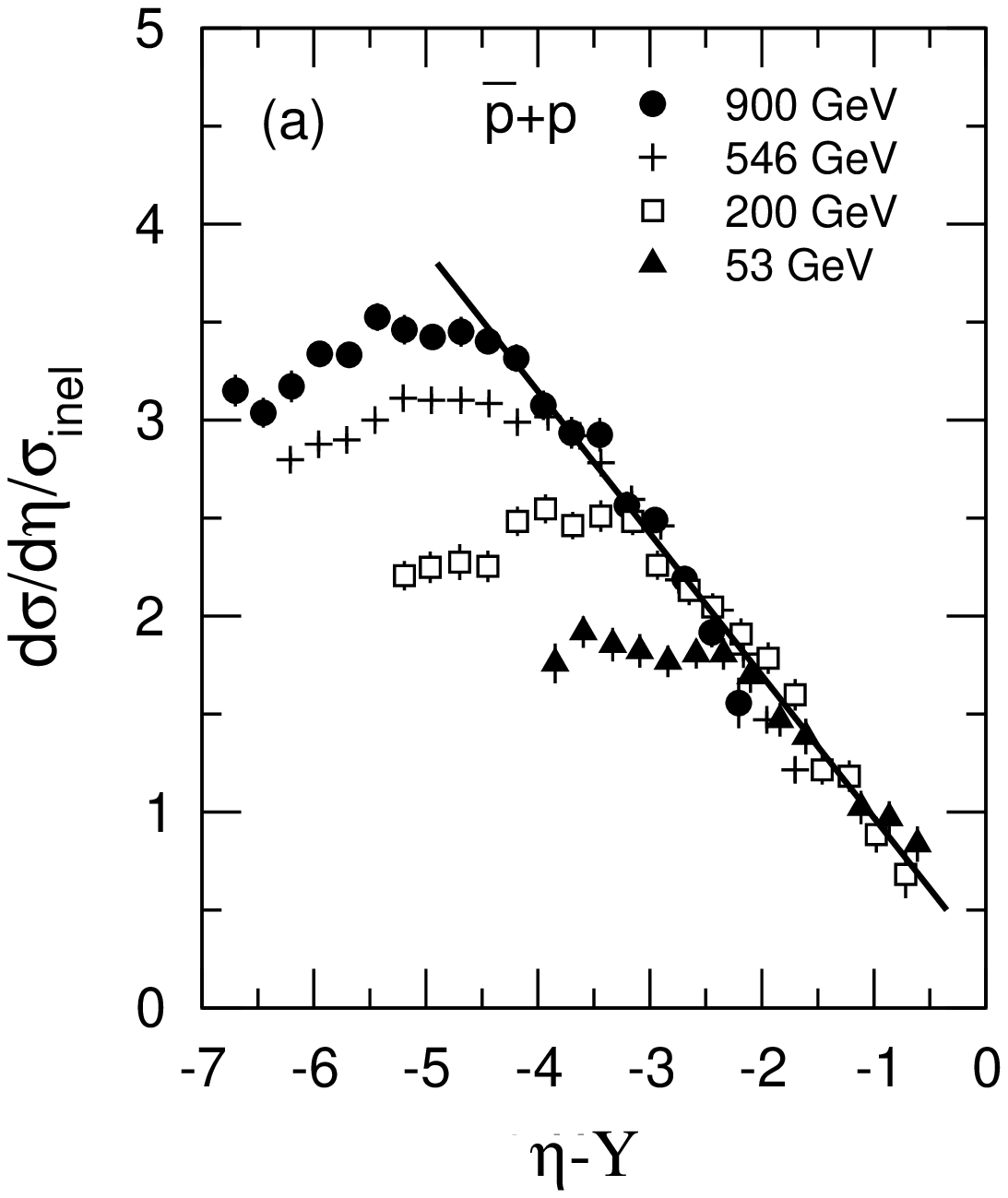,width=6cm}
\epsfig{file=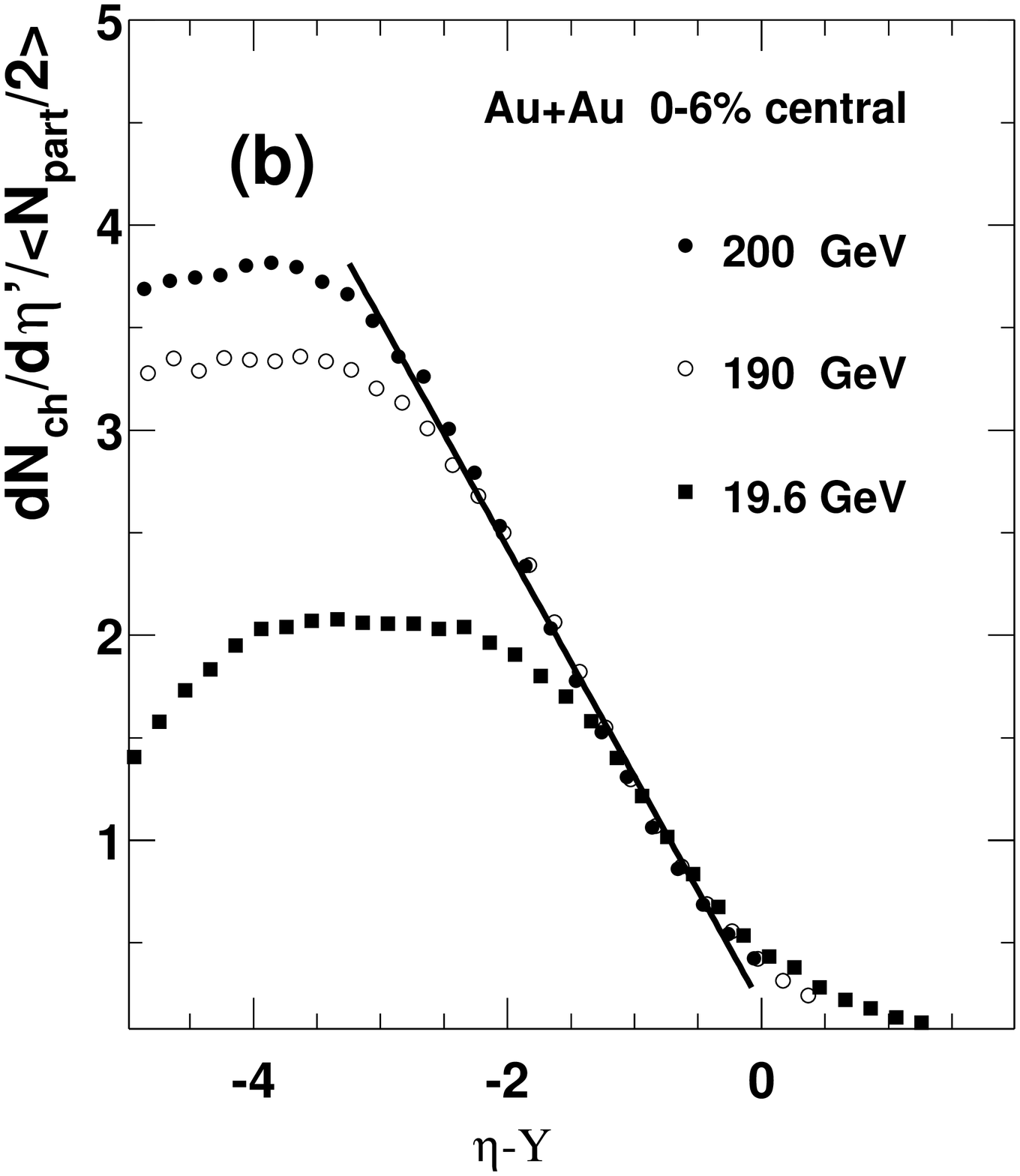,width=6cm}}
\caption{Particle density in pseudo-rapidity plotted versus 
$Y-\eta$. (a) $p-\bar p$ collisions \cite{ua5}. (b) nucleus-nucleus collisions \cite{ph2}. 
Lines are drawn to guide the eye.}
\end{figure}

The phenomenon is illustrated in Fig. 1, taken from \cite{ua5}  (1a)
and \cite{ph1} (1b), where particle density in pseudo-rapidity is plotted
versus the difference between the beam rapidity $Y$ and pseudorapidity
$\eta$ of the particle. The data of UA5 collaboration \cite{ua5} on
$p\bar{p}$ collisions and recent data from PHOBOS collaboration
\cite{ph2} on Au-Au collisions are shown. One sees three prominent
features, common for the two data sets: (i) Except at very small
$Y-\eta$, particle density in the fragmentation region increases {\it
linearly} with increasing $Y-\eta$; (ii) This linear increase is
followed by a "plateau" in the central rapidity region; (iii) The width
of the central plateau grows with increasing $Y$ only very slowly, if at
all. This last feature implies that with inreasing $Y$ (i.e. increasing
energy) the range of the limiting fragmentation region increases
proportionally to $Y$.

\newpage
It should be emphasized that these features are in blatant disagreement
with the principle of boost-invariance in particle production \cite{f}.
They are thus difficult to understand in the standard description of
production processes where one expects the particle density to be
dominated by the central plateau \cite{f,bj}, as is common in many
current models. In contrast, as is seen in the Fig. 1, the "central
plateau" occupies only a fraction of the available rapidity range.

In the present note we show that all these features can be understood in
a picture where particle production proceeds by a number of colour
exchanges between the two sets of partons one from the projectile and
one from the target. These colour exchanges lead to creation of the
colour charges which emit the observed particles by the bremsstrahlung
process. If the original partons in each of the colliding hadrons are
uniformly distributed in rapidity (i.e. if they satisfy the Feynman
$dx/x$ rule) the resulting distribution of observed particles is linear
in $Y-y$. Noting that only partons with life-time longer than the time
$\tau_0$, needed for the colour exchange  to take place, can participate in
the process, we conclude that this linear increase of the spectrum must
stop at a rapidity $y=y_0$, depending on the parton transverse mass
$\mu$ and on $\tau_0$. Thus the linear increase is followed by a plateau
for $y$ smaller than $y_0$. By {\it postulating} that this picture is valid in
the c.m. frame of the collision, thus violating explicitely the
boost-invariance, one accounts for the gross features of the data.

In Sections 2 and 3 a more detailed description of the model is
presented. In Sections 4 and 5 the consequences of the model for
particle spectrum are described. Discussion and conclusions are given in
the last section.

{\bf 2.} We follow the standard approach to multiparticle production
thus accepting that high-energy collision of two hadrons can be
described by colour exchange between the partons from the projectile and
from the target \cite{low}\footnote {There are many models of this type
 and many reviews. See, e.g. \cite{du, mod}.}. 

The new idea which we propose in this note is that this mechanism is
realized by colour exchange between several pairs of partons chosen at
random, one from the projectile and one from the target. The members of
each pair moving in opposite direction, each one radiates the observed
particles (or particle clusters\footnote{ E.g.in the form of hadronic
resonances.}) in the process of bremsstrahlung \cite{sto}. Note that by
this postulate we abandon the Feynman assumption \cite{f} that the
interaction is dominated by "wee" partons, i.e. by partons with small
rapidities.

To justify this idea we observe that it can be realized simply by one
gluon exchange process. Indeed, the gluon being particle of spin one,
its exchange gives the energy-independent cross-section. This means that
the probability of interaction between two partons by exchanging a gluon
is independent of their relative rapidity. In short, our assumption is
justified by the existence of  vector particles in QCD.

To obtain specific predictions one needs to know the shape of the parton
distribution. We simply accept the standard idea that the partons in
each of the colliding hadrons are distributed uniformly in rapidity
 and we can thus write their distribution in the form \cite{sto} 
\ba
dn(z_+)=  b(1-z_+)^{b-1} \frac{dz_+}{z_+}      \lab{1}
\ea
where $b$ is the parton density per unit of rapidity and
$z_+=(E+p_L)/(E_i+P_i)$, with $(E,p_L)$ are the energy and longitudinal
momentum of the parton, whereas $(E_i,P_i)$ are the energy and momentum of
the beam\footnote{These formulae apply for the right-moving system. For
the left-movers one should replace $z_+$ by $z_-=(E-p_L)/(E_i-P_i)$.}.

This picture makes sense only if the partons in one projectile can be
treated as independent from those in the other one. This
may be justified  if the
rapidity separation is large enough. When the rapidity separation is
small, however, (i.e. in the region close to the rapidity of the center
of mass) the partons from the two projectiles mix up and one cannot
expect them to act independently. Also the notion of the colour
separation  looses the meaning. This case thus demands a special
attention. 

This "very central" region is best studied in the overall c.m. frame. In
this frame the parton energies are not large and thus their distribution
rapidly fluctuates in time. To participate in the collision process,
however, the life-time of a parton must be substantially longer than the
time $\tau_0$ needed for the interaction to take place. This condition
allows to estimate the effective size of the rapidity region which does not
contribute to the particle production.

To see this we observe that the life-time of a high-energy
 parton with transverse mass equal to $\mu$
 can be estimated from  the uncertainty principle as
\ba
\tau\approx \gamma/\mu = E/\mu^2     \lab{3}
\ea
where $\gamma$ is the Lorentz factor.
From the condition $\tau \gg \tau_0 $  we  obtain $\mu e^{y'}/2\mu^2 \gg
\tau_0 $ where $y'$ is the rapidity of the parton. This implies
\ba
e^{y'}\geq e^{y_0}\gg 2\mu \tau_0 ;\;\;\;\; z \geq z_0 \gg \frac{\mu^2\tau_0}{E_i+P_i}
      \lab{4}
\ea

One sees that the condition (\ref{4}) restricts substantially the
rapidity of partons which can participate in particle production, as one
may expect $\tau_0$ to be of the order of 1 fermi($\tau_0\approx 1/p_t$,
where $p_t$ is the transverse momentum exchanged in the
interaction)\footnote{This restriction is of course much less effective
for {\it hard} collisions where the interaction time may be very
short.}.

{\bf 3.} The emission of particle clusters in the bremsstrahlung process
was analyzed some time ago by Stodolsky \cite{sto}. We follow his approach and
write the  particle distribution in the form
\ba
dN(x_+) = a (1-x_+)^{a-1} \frac{dx_+}{x_+}      \lab{a1}
\ea
where $a$ is the density of emitted hadrons per unit of rapidity
 and $x_{\pm}= (\epsilon\pm q_L)/(E_i+P_i)$, with 
$(\epsilon, q_L)$ being the energy and longitudinal momentum of the
emitted cluster.

Denoting by 
$\lambda$  the fraction of  "active" partons, i.e., the partons which participated in
the collision and using  (\ref{1}), the
distribution of the bremsstrahlung products is
\ba
dN(x_+)= \lambda \int_{\hat{z}}^1 b(1-z_+)^{b-1}\frac{dz_+}{z_+}
\left[a\left(1-\frac{x_+}{z_+}\right)^{a-1}\frac{dx_+}{x_+}\right] \lab{a2}
\ea
where the lower limit of integration $\hat{z}$ is 
\ba
\hat{z} = max(x_+,z_0)   \lab{a2a}
\ea
with $z_0$ determined by the energy below which a parton does not live
long enough to undergo a soft interaction and therefore also does not
radiate(c.f. (\ref{4}) and the related discussion in the previous
section).

By changing the variables:
\ba
u=1-x_+;\;\;\;\hat{u} =1-\hat{z};\;\;\;z_+=1-ut   \lab{a4a}
\ea
we obtain 
\ba
x_+\frac {dN}{dx_+}\equiv \frac{dN}{dy}=\lambda
 abu^{a+b-1}\int_0^{\hat{u}/u} dt(1-ut)^{-a}(1-t)^{a-1}t^{b-1} \lab{a5}
\ea

{\bf 4.} We are mostly interested in the particle distribution for
rapidities outside the projectile fragmentation region, i.e, for small
$x_+ \approx 0$. The formula (\ref{a5}) shows that we have to consider
two cases.

For  $x_+<z_0$ we  have $\hat{u} = 1-z_0 $. Thus, in the limit $x_+\rightarrow 0$
we obtain 
\ba
\frac {dN}{dy}\;\;\rightarrow\;\; \lambda ab \int_0^{u_0} \frac{dt}{1-t}t^{a-1}
\lab{a6a}
\ea
One sees that the result is independent of $u=1-x_+$, i.e. we obtain a plateau for
$y\leq y_0$.

For $x \geq z_0$ we  have $\hat{z}=x_+$, i.e. $\hat{u}=u$.
 Consequently, (\ref{a5}) can be rewritten as
 
\ba
\frac{dN}{dy}=\lambda a b \frac{\Gamma(a)\Gamma(b)}{\Gamma(a+b)}
u^{a+b-1} F(a,b;a+b;u)  \lab{a6}
\ea
where $F$ is the hypergeometric function. Note that the result is
perfectly symmetric with respect to $a$ and $b$. 

To see the behaviour at $x_+\approx 0$, i.e. $u\approx 1$ we use the
formula giving expansion of $F(a,b;a+b;u)$ around $u=1$ \cite{abr}. In the limit
of small $x_+$ this gives
\ba
\frac{dN}{dy}=  \lambda ab
\left[2\psi(j+1)-\psi(a+j)-\psi(b+j)-\log x_+\right]=\n=
 \lambda ab
\left[2\psi(1)-\psi(a)-\psi(b)+\log(M/m)+Y-y \right]   \lab{a13}
\ea
where  $m$ is the
transverse mass  of the emitted cluster and 
$M$ is the mass of the incident particle. We thus obtain a
linear increase with increasing $Y-y$, as observed in the data. 

The behaviour in the fragmentation region $x_+\approx 1$,  $u\approx 0$, is best seen
from(\ref{a6}). The result is
\ba
\frac {dN}{dy}=  \lambda\frac{\Gamma(a+1)\Gamma(b+1)}{\Gamma(a+b)}
(Y-y)^{a+b-1}  \lab{a14}
\ea
which shows a deviation from the linear increase unless $a+b=2$.

{\bf 5.} The distribution discussed in Section 4 can be justified for
positive rapidities but is not applicable in the negative rapidity
region. This is the standard problem in the bremsstrahlung model
\cite{sto}. The reason is clear: Eq (\ref{a1}) was obtained by requiring
conservation of the sum $(\epsilon+q_L)$, ignoring entirely conservation
of the difference $(\epsilon-q_L)$ (i.e. ignoring entirely the target).

Since the division into positive and negative rapidities is
frame-dependent, the region of applicability of (\ref{a5}) is also
frame-dependent. To fix this, we are again forced to use the hypothesis
that our considerations are  valid in the c.m. frame of the
collision. Of course any other frame boosted by less than $y_0$ is
equally good.

In the actual calculations (shown in Fig. 2) we cut the distribution for
negative c.m. rapidities using
a simple
prescription to multiply the distribution (\ref{a5}) by the correcting
factor 
\ba \Phi_+(y)= \frac{x_+}{x_++x_-}= \frac{e^{2y}}{e^{2y} +1}
\lab{a15} \ea
 One sees that for positive (large) $y$ the correction factor
is unimportant. On the other hand, it cuts exponentially the
distribution for negative $y$. 
A similar procedure must be, of course, applied also to the other
projectile, where one takes  $\Phi_-(y)= \Phi_+(-y)=1-\Phi_+(y)$.
The observed distribution is the sum of two contributions, one from the
right-moving system and another one from the system moving to the left.

Since the model predicts a plateau in  rapidity between $-y_0$ and
$+y_0$, the exact form of the cut-off is not essential, as long as it is
ineffective beyond the plateau region \cite{sto}.

%rys.2
\begin{figure}[htb]
\centerline{%
\epsfig{file=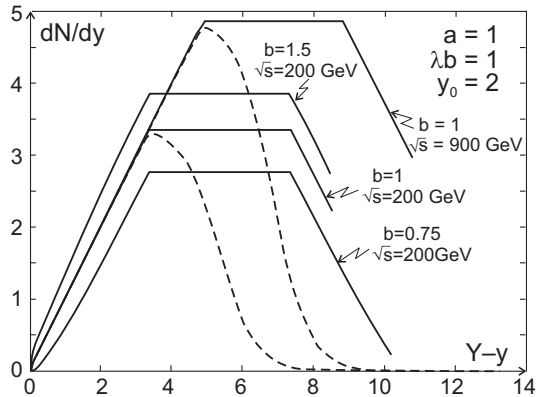,width=7cm}}
\caption{Particle density calculated from (\ref{a5}). Parameters as shown in the figure.
 The dashed lines show the effect of the cut-off (\ref{a15})}
\end{figure}

In Fig. 2 the calculated $dN/dy$ is plotted versus $Y-y$ for $a=1$,
$\lambda b=1$, $y_0=2$, $M=m$, two energies and three values of the parameter
$b$. One sees that the numerical results confirm the semi-quantitative
conclusions given in the previous section. One also sees that they
resemble nicely the data shown in Fig. 1.

{\bf 6.} Several comments are in order.

(i) The idea that the colour charges created in the first step of the
collision are responsible for particle production is rather general and
may be implemented in many ways. An interesting possibility is to
consider a more detailed model assuming that the active (radiating)
partons are colour octets and the density of hadronic clusters is
proportional to the local density of the chromoelectric field created by
these partons. For a given rapidity $\bar y$ there are, say, $n_L$
octets moving to the left (i.e. having rapidities smaller than $\bar y$)
and $n_R$ colour octets moving to the right (i.e. those with rapidities
greater than $\bar y$). Colour conservation implies that the
representation $R_L$ formed by left movers is conjugate to the
representation $R_R$ formed by right movers: $R_R = {\bar{R_L}}$

If the local energy density ${\cal E}(y)$ of the 
chromoelectric field is proportional to the quadratic
Casimir operator 
$C_2\left( R_L\right) = C_2\left( R_R\right)$
and for $n_L \le n_R$
\begin{equation}
R_L = 8 \otimes 8\otimes \dots 8 \qquad (n_L {\rm times}) 
\end{equation}
then the average energy density
\begin{equation}
\left< {\cal E} \right>_{n_L,n_r} = 
{\cal E}_0 C_A min\left(n_L, n_R\right)
\end{equation}
where $C_A = 3$ is the value of the Casimir operator for
the adjoint representation and ${\cal E}_0$ is a constant.
The linear increase of hadron rapidity  density is obtained 
if the rapidity distribution of radiating partons
is uniform in both  hemispheres.  The maximum of the single particle
distribution is obtained for the rapidity corresponding to $n_L=n_R$.
In $p-p$ collisions this is of course the rapidity of the center of
mass.

(ii) We have worked out in detail the hypothesis that the production of
final hadrons from the colour charges proceeds by the bremsstrahlung
process. This formulation is by no means unique. Conclusions similar to
the ones reached in this paper can be obtained if the production is
described by breaking of colour strings spanned between a parton from
the projectile and a parton from the target. In this case the cut-off
(\ref{a15}) is not needed. Instead, to obtain the plateau in the central
region one may postulate that the difference between the ends of a
string  contributing to particle production must exceed $2y_0$. This
does not change the main results of the model. The essential point,
needed to obtain the linear increase of the rapidity spectrum, is the
flat distribution (in rapidity) of the radiating partons and the flat
distribution of clusters in string decay (with energy independent
density). 

One should note, however, that these two versions of the model
give observable differences for asymmetric heavy ion collisions (in
particular p-A and d-A collisions) and for forward-backward correlations
in particle multiplicities. It may thus be interesting to study these
correlations experimentally.

(iii) The model contains several unknown parameters which, fortunately,
have a well-defined physical meaning. Their determination from the data
may thus give an interesting insight into structure of hadrons {\it
relevant for soft interactions}. For example, a determination of the
effective parton density at low momentum transfers (described by the
parameters $b$ and $\lambda$) is of clear interest. When applied to
nuclear collisions, this would allow to investigate the relation between
the effective parton densities and other parameters such as the number
of wounded nucleons \cite{bbc} and/or number of collisions.

(iv) Our argument explains only the gross features of the data. To
obtain a more detailed description, it is necessary to include at least
the effects of cluster decays\footnote{Also the change of variables ($y
\leftrightarrow \eta$) may be important, particularly in the region $y
\approx Y$ \cite{gbf}.}. This seems feasible, particularly in
pseudorapidity, where the isotropic clusters have a well-known decay
distribution.

{\bf 7.} In conclusion, to explain the observed strong violation of
boost invariance in rapidity spectra, we have proposed a two-step
mechanism for soft particle production in hadronic collisions. The first
step is the multiple gluon exchange between the partons from the two
colliding hadrons. In the second step, partons which were involved in
this process radiate hadronic clusters. This mechanism provides a
natural explanation of the observed rapidity spectra, in particular
their {\it linear} increase with increasing rapidity distance from the
maximal rapidity. Also the short plateau in the central rapidity region
is naturally obtained. The scheme can be applied also to nuclear
collisions and is flexible enough to account for the gross features of
the data. All parameters needed in this description have a well-defined
physical meaning and thus their determination from the data would give
useful information on hadron structure in the non-perturbative region.

\vspace{0.3cm}

 {\bf Acknowledgements}

\vspace{0.3cm}

Thanks are due to K.Fialkowski, W.Florkowski, K.Golec-Biernat,
R.Holynski, B.Wosiek and K.Wozniak for illuminating discussions and for
help in preparing the manuscript. A.B. thanks L. Mc Lerran for a comment
which directed his interest to the problem treated in the present note.
This investigation was supported in part by the by the Polish State
Commitee for Scientific research (KBN) Grant No 2 P03 B 09322 and 2 P03
B 00122.

\end{document}